\documentclass[12pt,psfig,preprint,letterpaper]{article}
\usepackage[centertags]{amsmath}
\usepackage{amssymb}
\newcommand{\Z}{{\mathbb Z}}

\begin{document}

\begin{flushright}
\baselineskip=12pt \normalsize
{ACT-11-05}\\
{MIFP-05-31}\\
\smallskip
\end{flushright}

\begin{center}
\Large {\textbf{Flipped $SU(5)$ From D-branes With Type IIB Fluxes}}        \\[2cm]
\normalsize Ching-Ming Chen$^{\dag\,1}$,  V. E. Mayes$^{\dag\,2}$,
D. V. Nanopoulos$^{\dag\, \ddag\, \S\,3}$
\\[.25in]
\textit{$^\dag$George P. and Cynthia W. Mitchell Institute for
Fundamental Physics, \\Texas A$\&$M University, College Station,
TX 77843, USA
\\
$^\ddag$Astroparticle Physics Group, Houston Advanced Research
Center (HARC), \\Mitchell Campus, Woodlands, TX 77381, USA
\\
$^\S$Academy of Athens, Division of Natural Sciences, \\28
Panepistimiou Avenue, Athens 10679, Greece} \\[.5cm]
\tt \footnotesize $^1$cchen@physics.tamu.edu,
$^2$eric@physics.tamu.edu,\\
$^3$dimitri@physics.tamu.edu\\[2cm]
\end{center}

\begin{abstract}
We construct flipped $SU(5)$ GUT models as Type IIB flux vacua on
$\Z_2\times \Z_2$ orientifolds. Turning on supergravity self-dual
NSNS and RR three-form fluxes fixes the toroidal complex structure
moduli and the dilaton.  We give a specific example of a
three-generation flipped $SU(5)$ model with a complete Higgs
sector where supersymmetry is softly broken by the supergravity
fluxes in the closed string sector.  All of the required Yukawa
couplings are present if global $U(1)$ factors resulting from a
generalized Green-Schwarz mechanism are broken spontaneously or by
world-sheet instantons.  In addition, the model contains extra
chiral and vector-like matter, potentially of mass
$\mathcal{O}(M_{string})$ via trilinear superpotential couplings.
\end{abstract}

\newpage
\setcounter{page}{1}
\pagestyle{plain}

\section{Introduction}

The fundamental goal of string phenomenology is to find a
convincing connection between realistic particle physics and
string theory. Previously it was thought that only models based
upon weakly coupled heterotic string compactifications could
achieve this.  Indeed, the most realistic model based on string
theory may be the heterotic string-derived flipped
$SU(5)$~\cite{AEHN} which has been studied in great detail.
However, in recent years Type I and Type II compactifications
involving D-branes, where chiral fermions can arise from strings
stretching between D-branes intersecting at angles (Type IIA
picture) \cite{Berkooz:1996km} and in its T-dual (Type IIB)
picture with magnetized D-branes \cite{Bachas:1995ik}, have
provided an interesting and exciting approach to this problem.

Many consistent standard-like and grand unified theory (GUT)
models were built at an early stage \cite{Krause:2000gp,
Blumenhagen:2000wh, Aldazabal:2000dg, Angelantonj:2000hi, List}
using D-brane constructions.  However, these models encountered
problems of supersymmetry. Furthermore, these models suffered from
instability in the internal space. The quasi-realistic
supersymmetric models were constructed first in Type IIA theory on
a $\mathbf{T^6} /(\Z_2 \times \Z_2)$ orientifold
\cite{Forste:2000hx, Cvetic:2001tj, CveticShiuUranga} and other
orientifolds \cite{BGOH}. Following this, models with
standard-like, left-right symmetric (Pati-Salam), Georgi-Glashow
($SU(5)$) and flipped $SU(5)$ gauge groups have been constructed
based upon this framework and systematically studied
\cite{Cvetic:P:S:L, Dijkstra:2004ym, Cvetic:2004nk, Chen:2005ab}.

However, in spite of these successes, a natural mechanism is still
needed to stabilize the moduli of the compactification, although
in some cases the complex structure parameters (in Type IIA
picture) and dilaton fields may be stabilized due to the gaugino
condensation in the hidden sector \cite{CLW}. Turning on RR and
NSNS fluxes as background of the compactification gives rise to a
non-trivial low energy supergravity potential which freezes some
Calabi-Yau moduli \cite{GVW}.  Type IIB configurations with
non-trivial RR and NSNS fluxes together with the presence of
anti-D3 branes have been studied in \cite{GiddingsKP,
Kachru:2002he}.  These fluxes impose strong constraints on the RR
tadpole cancellation by giving large positive D3 RR charges
since their supergravity equation of motion and the Dirac
quantization conditions must be satisfied.

In the closed string sector, generic choices of the fluxes do not
preserve supersymmetry. This leads to soft supersymmetry breaking
terms at a mass scale $M_{soft}\sim \frac{M_{string}^2}{M_{Pl}}$
which implies an intermediate string scale or an inhomogeneous
warp factor in the internal space to stabilize the electroweak
scale \cite{Camara:2003Soft, Lust:2004fi, CveticLiLiu}. On the
other hand, for the string scale to be close to the Planck scale,
supersymmetry in the open string sector must be preserved by
fixing the K\"{a}hler toroidal moduli \cite{CveticLiLiu}, which is
T-dual to the supersymmetry consistency conditions in Type IIA
theory. Recently D-brane constructions corresponding to models
with magnetized D-branes where the role of the intersection angles
is played by the magnetic fluxes on the D-branes on the Type IIB
$\mathbf{T^6} /(\Z_2 \times \Z_2)$ orientifold have been
studied~\cite{ CveticLiLiu, Blumenhagen:2003vr, CascalesUranga,
MarchesanoShiu, CLN}.

As previously mentioned, there are only a few specific choices of
fluxes which are supersymmetric in the closed string sector, which
is interesting from a phenomenological point of view.  In general,
non-supersymmetric fluxes lead to soft supersymmetry breaking
terms in the effective action of open string fields. Detailed
studies of the soft-breaking mechanism and some trial
investigations in the effective low energy scenario were explored
in \cite{Lust:2004fi, Kane:2004hm}.  Combined with an analysis of
the Yukawa couplings \cite{Cremades:2003qj}, these studies may
provide a clear picture of the low energy physics in the
intersecting D-brane configuration which is worthwhile for future
work.  On the other hand, if supersymmetry is required to be
conserved both in the closed and open string sectors, it has been
recently shown that the RR, NSNS and metric fluxes could
contribute negative D6-brane charges in Type IIA orientifold with
flux compactifications, which makes it easier to satisfy the RR
tadpole cancellation conditions~\cite{Camara:2005dc}. We will
presently not consider this, but plan to investigate this
possibility in the future.

In this paper we search for consistent flipped $SU(5)$ models on a
Type IIB $\mathbf{T^6} /(\Z_2 \times \Z_2)$ orientifold with
supergravity fluxes turned on.  As mentioned before, due to the
difficulty to impose supersymmetric fluxes in the closed string
sector with consistent RR tadpole conditions, we do not insist
that the fluxes be supersymmetric and consider all possible fluxes
in constructing flipped $SU(5)$ models. However, supersymmetry in
the open string sector is still preserved for a reasonable string
scale.  By requiring that the gauge bosons coupled to $U(1)_X$ do
not acquire a string scale mass via a generalized Green-Schwarz
mechanism, which has four constraints in $(\Z_2 \times \Z_2)$
orientifold construction, we find that the models must have at
least five stacks of D-branes. In addition, there are K-theory
constraints which must be imposed to avoid the anomaly classified
by the discrete symmetry $\Z_2$. Some K-theory properties are
modified by the NSNS fluxes, but this is currently regarded to
have no effect on phenomenology \cite{CascalesUranga}.

Next, we turn to the question of our motivation in building
flipped $SU(5)$ models. Different types of particle models have
been discussed using various constructions.  The minimal option is
to embed just the Standard Model $SU(3)\times SU(2)\times U(1)$
gauge group, but almost every construction contains at least some
extra $U(1)$ factors. Conventional GUT models such as $SU(5)$ or
$SO(10)$ have been investigated, but none of them has been
completely satisfactory.  This triggered the motivation to
consider the gauge group $SU(5)\times U(1)_X$ \cite{AEHN,
Barr:1981qv, FSU(5)N} as a candidate for a model derived from
string.  The $raison\ d'\hat{e}tre$ of this \lq flipped\rq~$SU(5)$
is that it requires only $\mathbf{10}$ and
$\overline{\mathbf{10}}$ Higgs representations to break the GUT
symmetry, in contrast to other unified models which require large
and unwieldy adjoint representations.  This point was given
further weight when it was realized that models with adjoint Higgs
representations cannot be derived from string theory with a $k =
1$ Kac-Moody algebra~\cite{ELN1}. There are many attractive
features of flipped $SU(5)$. For example, the hierarchy problem
between the electroweak Higgs doublets and the color Higgs
triplets is solved naturally through a \lq missing
partner\rq~mechanism \cite{AEHN}. Furthermore, this dynamical
doublet-triplet splitting does not require or involve any mixing
between the Higgs triplets leading to a natural suppression of
dimension 5 operators that may mediate rapid proton decay and for
this reason it is probably the simplest GUT to survive the
experimental limits placed upon proton lifetime
\cite{Ellis:2002vk}.  Recent investigation showed that the proton
could be even stable by rotating away the gauge dimension 6
contributions \cite{Dorsner:2004jj}.  More recently, the cosmic
microwave anisotropy $\delta T/T$ has been successfully predicted
by flipped $SU(5)$, as it has been determined to be proportional
to $(M/M_P)^2$ where $M$ denotes the symmetry breaking scale and
$M_P = 2.4\times 10^{18}$~GeV is the reduced Planck mass
\cite{Kyae:2005nv}. Finally, string-derived flipped $SU(5)$ may
provide a natural explanation for the production of Ultra-High
Energy Cosmic Rays (UHECRs), through the decay of super-heavy
particles dubbed \lq cryptons\rq~\cite{cryptons}~ that arise in
the hidden sector of the model, which are also candidates for
cold-dark matter (CDM).

The heterotic string-derived flipped $SU(5)$ model was created within the
context of the free-fermionic formulation, which easily yields
string theories in four dimensions.  This model belongs to a class
of models that correspond to compactification on the $\Z_2\times
\Z_2$ orbifold at the maximally symmetric point in the Narain
moduli space~\cite{Narain}. Although formulated in the context of
weakly coupled heterotic string theory, it is believed that the
vacuum may in fact be non-perturbative due to it's proximity to
special points in the moduli space and may elevate to a consistent
vacuum of M-theory.  For this reason, it is our
hope that in searching for a realistic flipped $SU(5)$ model
that we may arrive at or near the same vacuum using
D-brane constructions.

We organize this letter in the following way.  In section 2 a
brief but complete construction of D-branes compactified on
$\mathbf{T^6} /(\Z_2 \times \Z_2)$ with Type IIB RR and NSNS
supergravity fluxes is provided.  In section 3 a short
review of basic flipped $SU(5)$ phenomenology is presented.
Section 4 contains the discussion of D-brane model building with
fluxes, and we provide a few examples including a complete
spectrum of a flipped $SU(5)$ model. We present our conclusions
in section 5.

\section{D-branes with Type IIB Flux on the $\mathbf{T^6} /(\Z_2
\times \Z_2)$ Orientifold}

\subsection{Magnetized D-branes in Type IIB Theory}

We begin with the Type IIB theory on the $\mathbf{T^6} /(\Z_2 \times
\Z_2)$ orientifold, where $\mathbf{T^6}$ is product of three
two-tori and the two orbifold group generators $\theta$, $\omega$
act on the complex coordinates $(z_1,z_2,z_3)$ as
\begin{eqnarray}
\theta:(z_1,z_2,z_3)\rightarrow(-z_1,-z_2,z_3) \nonumber \\
\omega:(z_1,z_2,z_3)\rightarrow(z_1,-z_2,-z_3)
\end{eqnarray}

This construction contains a $D=4$, $N=2$ supergravity multiplet,
the dilaton hypermultiplet, $h_{11}$ hypermultiplets, and $h_{21}$
vector multiplets which are all massless.  For the orbifold with
discrete torsion the Hodge numbers from both twisted and untwisted
sectors are $(h_{11},h_{21})=(3,51)$.  In order to include the
open string sector, orientifold planes are introduced by an
orientifold projection $\Omega R$, where $\Omega$ is the
world-sheet parity and $R$ acts as
\begin{equation}
R:(z_1,z_2,z_3)\rightarrow(-z_1,-z_2,-z_3)
\end{equation}

There will then be 64 $O3$-planes and 4 $O7_i$-planes, which are
transverse to the $\mathbf{T}^2_i$.  Thus $\Omega R$ projects the
$N=2$ spectrum to an $N=1$ supergravity multiplet, the dilaton
chiral multiplet, and 6 untwisted and 48 twisted geometrical
chiral multiplets. \cite{CveticLiLiu, MarchesanoShiu}

We need $D(3+2n)$-branes to fill up the four-dimensional Minkowski
space-time and wrapping the $2n$-cycles on a compact manifold in
type IIB theory.  The introduction of magnetic fluxes provides
more flexibility in constructing models.  For one stack of $N_a$
D-branes wrapping $m^i_a$ times on $\mathbf{T}^2_i$, $n^i_a$
denotes the units of magnetic fluxes $F^i_a$ turned on each
$\mathbf{T}^2_i$, thus
\begin{equation}
m^i_a \frac{1}{2\pi}\int_{\mathbf{T}^2_i} F^i_a = n^i_a
\end{equation}

To write down an explicit description of D-brane topology we
introduce the even homology classes $[\mathbf{0}_i]$ and
$[\mathbf{T}_i]$ for the point and the two-torus.  Then the
vectors of RR charges (corresponding to Type IIA homology cycles)
of $a^{th}$ stack D-brane and its image are \cite{CascalesUranga}
\begin{equation}
[\Pi_a]=\prod^3_i (n^i_a [\mathbf{0}_i]+m^i_a [\mathbf{T}_i]), \;
[\Pi_a']=\prod^3_i (n^i_a [\mathbf{0}_i]-m^i_a [\mathbf{T}_i])
\end{equation}

The O3- and $\mathrm{O7}_i$-planes of $\mathbf{T^6} /(\Z_2 \times
\Z_2)$ resulting from the orientifold action $\Omega R$, $\Omega R
\omega$, $\Omega R \theta \omega$ and $\Omega R \theta$ can be
written as
\begin{eqnarray}
\Omega R :& [\Pi_{O3}]= [\mathbf{0}_1][\mathbf{0}_2][\mathbf{0}_3]
\nonumber \\
\Omega R \omega :& [\Pi_{O7_1}]= -
[\mathbf{0}_1][\mathbf{T}^2_2][\mathbf{T}^2_3] \nonumber \\
\Omega R \theta \omega :& [\Pi_{O7_2}]= -
[\mathbf{T}^2_1][\mathbf{0}_2][\mathbf{T}^2_3] \nonumber \\
\Omega R \theta :& [\Pi_{O7_3}]= -
[\mathbf{T}^2_1][\mathbf{T}^2_2][\mathbf{0}_3] \label{O-planes}
\end{eqnarray}
where the total effect is the sum of the above O-planes:
$[\Pi_{O_p}]=[\Pi_{O3}]+[\Pi_{O7_1}]+[\Pi_{O7_2}]+[\Pi_{O7_3}]$.

\subsection{The Fermionic Spectrum}

\begin{table}[h]
\renewcommand{\arraystretch}{1.5}
\center
\begin{tabular}{|c||c|}
\hline

Sector & Representation   \\ \hline \hline

$aa$ & $U(N_a /2)$ vector multiplet and 3 adjoint chiral multiplets \\
\hline

$ab+ba$ & $ {\cal M}(\frac{N_a}{2},
\frac{\overline{N_b}}{2})=I_{ab}=\prod_{i=1}^3(n^i_a m^i_b - n^i_b
m^i_a) $  \\ \hline

$ab'+b'a$ & $ {\cal M}(\frac{N_a}{2},
\frac{N_b}{2})=I_{ab'}=-\prod_{i=1}^3(n^i_a m^i_b + n^i_b m^i_a) $
\\ \hline

$aa'+a'a$ & $  {\cal M}({\rm
Anti}_a)=\frac{1}{2}(I_{aa'}+\frac{1}{2}I_{aO})

 $ \\
 & $ {\cal M}({\rm
Sym}_a)=\frac{1}{2}(I_{aa'}-\frac{1}{2}I_{aO})$
\\ \hline

\end{tabular}
\caption{Spectrum of bi-fundamental representations, where
$I_{aa'}=-8\prod^3_{i=1}n^i_a m^i_a$, and $I_{aO}=8(-m^1_a m^2_a
m^3_a + m^1_a n^2_a n^3_a +n^1_a m^2_a n^3_a + n^1_a n^2_a
m^3_a)$.}
\end{table}

Chiral matter arises from open strings with two ends attaching on
different stacks.  The multiplicity (${\cal M}$) of the
corresponding bi-fundamental representation is given by the \lq
intersection\rq~number (as in Type IIA theory) between different
stacks of branes.  The initial $U(N_a)$ gauge group supported by a
stack of $N_a$ identical D6-branes is broken down by the
$\Z_2\times \Z_2$ symmetry to a subgroup $U(N_a/2)$. However a
model may contain additional non-chiral (vector-like) multiplet
pairs from $ab+ba$, $ab'+b'a$, and $aa'+a'a$ if the branes are
parallel on at least one torus. The multiplicity of these
non-chiral multiplet pairs is given by the remainder of the
intersection product, neglecting the null sector. For example, if
$(n^1_a m^1_b - n^1_b m^1_a)=0$ in $
I_{ab}=[\Pi_a][\Pi_b]=\prod_{i=1}^3(n^i_a m^i_b - n^i_b m^i_a) $,
\begin{equation}
{\cal M}\left[\left(\frac{N_a}{2},\frac{\overline{N_b}}{2}\right)
+\left(\frac{\overline{N_a}}{2},\frac{N_b}{2}\right)\right]
=\prod_{i=2}^3(n^i_a m^i_b - n^i_b m^i_a)
\end{equation}
The multiplicity of bi-fundamental as well as symmetric and
antisymmetric representations are shown in Table 1.

\subsection{Turning on Type IIB Fluxes}

Turning on supergravity fluxes for closed string fields provides a
possible way to stabilize the compactification moduli; however it
also naturally breaks space-time supersymmetry in the bulk as well
as contribute to the RR charges. Thus, specific solutions are needed
to preserve supersymmetry.

The Type IIB non-trivial RR 3-form $F_3$ and NSNS 3-form $H_3$ fluxes
compactified on Calabi-Yau threefold $X_6$ need to obey the Bianchi
identities and be quantized \cite{GiddingsKP}:
\begin{equation}
dF_3=0,\;\; dH_3=0
\end{equation}
\begin{equation}
\frac{1}{(2\pi)^2 \alpha'}\int_{X6}F_3 \in \mathbf{Z},\;\;
\frac{1}{(2\pi)^2 \alpha'}\int_{X6}H_3 \in \mathbf{Z}
\end{equation}

When the two fluxes are turned on, they induce a covariant field
$G_3=F_3-\tau H_3$ and contribute to the D3-brane RR charges
\begin{equation}
N_{flux}=\frac{1}{(4\pi^2 \alpha')^2}\int_{X_6} H_3 \wedge F_3 =
\frac{1}{(4\pi^2 \alpha')^2}\frac{i}{2\mathrm{Im}(\tau)}\int_{X_6}
G_3 \wedge \bar{G_3}
\end{equation}
where $\tau=a+i/g_s$ being the Type IIB axion-dilaton coupling.

A complex cohomology basis can be utilized to describe the 3-form
flux $G_3$ on $\mathbf{T^6} /(\Z_2 \times \Z_2)$:
\begin{eqnarray}
&&\omega_{B_0}=dz^1 \wedge dz^2 \wedge dz^3, \;\;
\omega_{A_1}=d\bar{z}^1 \wedge dz^2 \wedge dz^3, \nonumber \\
&&\omega_{B_1}=dz^1 \wedge d\bar{z}^2 \wedge d\bar{z}^3, \;\;
\omega_{A_2}=dz^1 \wedge d\bar{z}^2 \wedge dz^3, \nonumber \\
&&\omega_{B_2}=d\bar{z}^1 \wedge dz^2 \wedge d\bar{z}^3, \;\;
\omega_{A_3}=dz^1 \wedge dz^2 \wedge d\bar{z}^3, \nonumber \\
&&\omega_{B_3}=d\bar{z}^1 \wedge d\bar{z}^2 \wedge dz^3, \;\;
\omega_{A_0}=d\bar{z}^1 \wedge d\bar{z}^2 \wedge d\bar{z}^3
\end{eqnarray}
where $dz^i=dx^i+U_i dy^i$, $U_i$ are complex structure moduli.
Here $\omega_{B_0}$ corresponds to the (3,0) of the flux,
$\omega_{B_i}$ with $i=$1, 2, 3 correspond to (1,2) of the flux,
$\omega_{A_i}$ with $i=$1, 2, 3 correspond to (2,1), and
$\omega_{A_0}$ is (0,3) component of the flux.  Then the untwisted
3-form $G_3$ takes the form:
\begin{equation}
\frac{1}{(2\pi)^2 \alpha'}G_3=\sum_{i=0}^3(A^i \omega_{A_i} + B^i
\omega_{B_i})
\end{equation}
Therefore the contribution of the fluxes to the RR tadpole
condition $N_{flux}$ can be calculated in terms of the basis
defined above:
\begin{equation}
N_{flux} = \frac{1}{(4\pi^2 \alpha')^2}
\frac{i}{2\mathrm{Im}(\tau)} \int_{X_6} G_3 \wedge \bar{G_3} =
\frac{4\prod_{i=1}^3\mathrm{Im}(U^i)}{\mathrm{Im} (\tau)}
\sum_{j=0}^3 (|A^i|^2-|B^i|^2)
\end{equation}
The choice of fluxes may be positive
(ISD-fluxes\footnote{Imaginary self dual fluxes, lead to zero or
negative cosmological constant(to lowest order).}) or negative
(IASD-fluxes). However, in order to satisfy the supergravity
equation of motion, the BPS-like self-dual condition $\ast_6
G_3=iG_3$ demands $N_{flux}$ to be positive \cite{Kachru:2002he,
Lust:2004fi, CveticLiLiu}. The quantization conditions of $F_3$
and $H_3$ fluxes require that $N_{flux}$ be a multiple of 64.

\subsection{Supersymmetry Conditions}

$D=4$ $N=1$ supersymmetric vacua from flux compactification
require 1/4 supercharges of the ten-dimensional Type I theory be
preserved both in the open and closed string sectors
\cite{CveticLiLiu}.  The supersymmetry constraints in the open
string sector are from the world-volume magnetic field and those
in the closed string sector induced by the fluxes.

\subsubsection{Supersymmetry Conditions in the Closed String Sector}

In the closed string sector, to ensure that the RR and NSNS fluxes
are supersymmetric, the primitivity condition $G_3\wedge J =0$
should be satisfied \cite{Kachru:2002he}.  Here $J$ is the general
K\"{a}hler form of $\mathbf{T^6} /(\Z_2 \times \Z_2)$
\cite{CascalesUranga}:
\begin{equation}
J=J_1 dz^1 \wedge d\bar{z}^1 + J_2 dz^2 \wedge d\bar{z}^2 + J_3
dz^3 \wedge d\bar{z}^3
\end{equation}
We list a few solutions below.  We also require that the turned on fluxes are as
small as possible to avoid too large RR charge and satisfy the
above requirements.

\paragraph{(2, 1)-Flux}

\paragraph{(1)} A specific supersymmetric solution for $G_3$ is (2, 1)-form
given in \cite{Lust:2004fi} as
\begin{equation}
\frac{1}{(2\pi)^2 \alpha'}G_3= -4 \omega_{A_2} - 4\omega_{A_3}
\end{equation}
where the complex structure $U^i$ and the dilaton coupling $\tau$
stabilize at $U^1=U^2=U^3=\tau=i$.  This solution gives the flux
RR tadpole contribution:
\begin{equation}
N_{flux}=128
\end{equation}

\paragraph{(2)} Another specific supersymmetric solution for
(2, 1)-form is given in \cite{CascalesUranga} as
\begin{equation}
\frac{1}{(2\pi)^2 \alpha'}G_3= \frac{8}{\sqrt{3}} e^{-\pi
i/6}(\omega_{A_1} + \omega_{A_2} + \omega_{A_3})
\end{equation}
The fluxes stabilize the complex structure toroidal moduli at
values $U^1=U^2=U^3=\tau=e^{2\pi i/3}$.
Thus, the flux contributes to the RR tadpole contribution an amount:
\begin{equation}
N_{flux}=192
\end{equation}

\paragraph{Non-SUSY}
This solution has the smallest contribution to the D3 RR charge.
Although it is not supersymmetric due to the existence of (0, 3)
component, it is still worthy of study since we do not observe
supersymmetry at low energies. The 3-form flux is
\begin{equation}
\frac{1}{(2\pi)^2 \alpha'}G_3= 2( \omega_{A_0} + \omega_{A_1} +
\omega_{A_2} + \omega_{A_3})
\end{equation}
with $U^1=U^2=U^3=\tau=i$.  The flux induced RR charge is then
\begin{equation}
N_{flux}=64
\end{equation}

\subsubsection{Supersymmetry Conditions in the Open String Sector}

In order to preserve $N=1$ supersymmetry in the open string
sector, a constraint must be placed upon the D-brane world-volume
magnetic fields $F^i=n^i / m^i \chi^i$ associated with each
two-torus $\mathbf{T_i^2}$ which can be expressed in terms of an
\lq angle\rq~$\theta_i$ (as in the Type IIA picture) on each
torus, as $\sum_i \theta_i = 0$ mod $2\pi$ \cite{CascalesUranga},
where tan$\theta_i =(F^i)^{-1} = \frac{m^i \chi^i}{n^i}$ and
$\chi^i = R^i_1 R^i_2$ the area of the $\mathbf{T}^2_i$ in
$\alpha'$ units. Then we can write it in a form that is similar to
the constraints in Type IIA picture as \cite{CveticShiuUranga}
\begin{eqnarray}
-x_A m^1_a m^2_a m^3_a + x_B m^1_a n^2_a n^3_a + x_C n^1_a m^2_a
n^3_a + x_D n^1_a n^2_a m^3_a = 0 \nonumber \\
- n^1_a n^2_a n^3_a /x_A + n^1_a m^2_a m^3_a /x_B + m^1_a n^2_a
m^3_a /x_C + m^1_a m^2_a n^3_a /x_D < 0
\end{eqnarray}
where $x_A=\lambda$, $x_B=\lambda / \chi^2\chi^3$, $x_C=\lambda /
\chi^1\chi^3$, $x_D=\lambda / \chi^1\chi^2$, and $\lambda$ is a normalization
constant used
to keep the variables on an equal footing.

\subsection{RR Tadpole Cancellation and K-theory Constraints}

The RR charges of the magnetized D-brane associated homology
classes and the contribution from the orientifold planes as well
as the effect of the fluxes must be cancelled, namely we demand
\begin{equation}
\sum_a N_a [\Pi_a] + \sum_a N_a [\Pi_a'] + \sum_p N_{O_p} Q_{O_p}
[\Pi_{O_p}] + N_{flux} =0
\end{equation}
where [$\Pi_{O_p}$] are the sum of the orientifold planes listed
in (\ref{O-planes}), and $N_{O_p} Q_{O_p}=-32$ in
$\mathrm{D}_p$-branes for $S_p$-type O-planes. $N_{flux}$ is the
amount of flux turned on, and is quantized in units of the
elementary flux as discussed above \cite{Lust:2004fi, CveticLiLiu,
CascalesUranga}. Filler branes wrapping cycles along the O-planes
can also be introduced here to reduce the difficulty of satisfying
this condition.  Thus the RR tadpole cancellation condition can be
simplified as
\begin{eqnarray}
-N^{(O3)} - \sum_a N_a n^i_a n^2_a n^3_a - \frac{1}{2}N_{flux} =
-16 \nonumber \\
-N^{(O7_1)} + \sum_a N_a n^1_a m^2_a m^3_a =-16 \nonumber \\
-N^{(O7_2)} + \sum_a N_a m^1_a n^2_a m^3_a =-16 \nonumber \\
-N^{(O7_3)} + \sum_a N_a m^1_a m^2_a n^3_a =-16
\end{eqnarray}

In addition to the RR-tadpole condition the discrete D-brane RR
charges classified by $\mathbf{\Z_2}$ K-theory groups in the
presence of orientifolds, which are invisible by the ordinary
homology \cite{MarchesanoShiu, WittenK, UrangaK, Minasian:1997mm},
should be also taken into account \cite{MarchesanoShiu, UrangaK}.

In Type I superstring theory there exist non-BPS D-branes carrying
non-trivial K-theory $\mathbf{\Z_2}$ charges.  To avoid this
anomaly it is required that in compact spaces these non-BPS branes
must exist in an even number \cite{UrangaK}.  In Type IIB picture,
these Type I non-BPS $p$-branes can be regarded as a pair of
D$p$-brane and it's world-sheet parity image.  For example,
$\widehat{\textrm{D7}}|_\mathrm{I} = (\overline{\textrm{D}7} +
\overline{\textrm{D7}}/\Omega )|_{\mathrm{IIB}}$. We need to
consider the effects both from D3- and D7-branes since they do not
contribute to the standard RR charges. The K-theory conditions for
a $\mathbf{\Z_2\times \Z_2}$ orientifold were derived in
\cite{MarchesanoShiu} and are given by
\begin{eqnarray}
\sum_a N_a m^1_a m^2_a m^3_a = 0 \textrm{ mod }4,\;\; \sum_a N_a
m^1_a n^2_a n^3_a = 0 \textrm{
mod }4, \nonumber \\
\sum_a N_a n^1_a m^2_a n^3_a = 0 \textrm{ mod }4, \;\; \sum_a N_a
n^1_a n^2_a m^3_a = 0 \textrm{ mod }4. \, \label{K-charges}
\end{eqnarray}

Furthermore, D-brane states are classified by the K-theory group
due to the presence of NSNS 3-form fluxes as well. This requires
adding additional D-branes to preserve the homological charges and
the possibility of instanton  mediating D-branes and fluxes
\cite{CascalesUranga}.  These properties do not affect the main
constraints, and they are not presently well known and need
further study.

\subsection{The Green-Schwarz Mechanism for Flipped $SU(5)$ GUT Construction}

Although the total non-Abelian anomaly cancels automatically when
the RR-tadpole conditions are satisfied, additional mixed
anomalies like the mixed gravitational anomalies which generate
massive fields are not trivially zero \cite{CveticShiuUranga,
Cvetic-GS}.  These anomalies are cancelled by a generalized
Green-Schwarz (G-S) mechanism which involves untwisted
Ramond-Ramond forms.  The couplings of the four untwisted
Ramond-Ramond forms $B^i_2$ to the $U(1)$ field strength $F_a$ are
\cite{Aldazabal:2000dg}
\begin{eqnarray}
 N_a m^1_a n^2_a n^3_a \int_{M4}B^1_2\wedge \textrm{tr}F_a,  \;\;
 N_a n^1_a m^2_a n^3_a \int_{M4}B^2_2\wedge \textrm{tr}F_a
  \nonumber \\
 N_a n^1_a n^2_a m^3_a \int_{M4}B^3_2\wedge \textrm{tr}F_a,  \;\;
 -N_a m^1_a m^2_a m^3_a \int_{M4}B^4_2\wedge \textrm{tr}F_a
\end{eqnarray}
These couplings determine the linear combinations of $U(1)$ gauge
bosons that acquire string scale masses via the G-S mechanism. In
flipped $SU(5)\times U(1)_X$,  the symmetry $U(1)_X$ must remain a
gauge symmetry  so that it may remix to help generate the standard
model hypercharge after the breaking of $SU(5)$.  Therefore, we
must ensure that the gauge boson of the flipped $U(1)_X$ group
does not receive such a mass. The $U(1)_X$ is a linear combination
of the $U(1)$s from each stack :
\begin{equation}
U(1)_X=\sum_a c_a U(1)_a
\end{equation}
The corresponding field strength must be orthogonal to those that
acquire G-S mass.  Thus we demand :
\begin{eqnarray}
\sum_a c_a N_a m^1_a n^2_a n^3_a =0, \;\; \sum_a c_a N_a n^1_a
m^2_a n^3_a =0
  \nonumber \\
\sum_a c_a N_a n^1_a n^2_a m^3_a =0, \;\; \sum_a c_a N_a m^1_a
m^2_a m^3_a =0 \label{GSeq}
\end{eqnarray}

The G-S mechanism will be considered only after the coefficients
of $U(1)_X$ are determined.

\section{Flipped $SU(5)\times U(1)_X$ Model Building}

In the previous section we have outlined all the necessary
machinery for constructing models as Type IIB flux vacua on the
$T^6/(\Z_2\times \Z_2)$ orientifold.  Our goal now is to realize a
supersymmetric $SU(5)\times U(1)_X$ gauge theory with three
generations and a complete GUT and electroweak Higgs sector in the
four-dimensional spacetime. We also try to avoid as much extra
matter as possible.
\subsection{Basic Flipped $SU(5)$ Phenomenology}

In a flipped $SU(5)\times U(1)_X$ \cite{AEHN, Barr:1981qv,
FSU(5)N} unified model, the electric charge generator $Q$ is only
partially embedded in $SU(5)$, {\it i.e.}, $Q = T_3 -
\frac{1}{5}Y' + \frac{2}{5}\tilde{Y}$, where $Y'$ is the $U(1)$
internal $SU(5)$ and $\tilde{Y}$ is the external $U(1)_X$ factor.
Essentially, this means that the photon is \lq shared\rq \ between
$SU(5)$ and $U(1)_X$. The Standard Model (SM) plus right handed
neutrino states reside within the representations $\bar{\bf{5}}$,
$\bf{10}$, and $\bf{1}$ of $SU(5)$, which are collectively
equivalent to a spinor $\bf{16}$ of $SO(10)$.  The quark and
lepton assignments are flipped by $u^c_L$ $\leftrightarrow$
$d^c_L$ and $\nu^c_L$ $\leftrightarrow$ $e^c_L$ relative to a
conventional $SU(5)$ GUT embedding:
\begin{equation}
\bar{f}_{\bf{\bar{5},-\frac{3}{2}}}= \left( \begin{array}{c}
              u^c_1 \\ u^c_2 \\ u^c_3 \\ e \\ \nu_e
                    \end{array} \right) _L ; \;\;\;
F_{\bf{10,\frac{1}{2}}}= \left( \left( \begin{array}{c}
              u \\ d \end{array} \right) _L  d^c_L \;\; \nu^c_L
\right)
              ; \;\;\;
l_{\bf{1,\frac{5}{2}}}=e^c_L
\end{equation}
In particular this results in  the $\bf{10}$ containing a neutral
component with the quantum numbers of $\nu^c_L$.  We can
spontaneously break the GUT symmetry by using a $\bf{10}$ and
$\overline{\bf{10}}$ of superheavy Higgs where the neutral
components provide a large vacuum expectation value, $\left\langle
\nu^c_H \right\rangle$= $\left\langle \bar{\nu}^c_H
\right\rangle$,
\begin{equation}
H_{\bf{10,\frac{1}{2}}}=\left\{Q_H,\;d^c_H,\;\nu^c_H \right\};
\;\;\;
\bar{H}_{\bf{\overline{10},-\frac{1}{2}}}=\left\{Q_{\bar{H}},\;d^c_{\bar{H}},\;\nu^c_{\bar{H}}
\right\}.
\end{equation}
The electroweak spontaneous breaking is generated by the Higgs
doublets $H_2 $ and $ \bar{H}_{\bar{2}} $
\begin{equation}
h_{\bf{5,-1}}=\left\{ H_2,H_3 \right\}; \;\;\;
\bar{h}_{\bf{\bar{5},1}}=\left\{
\bar{H}_{\bar{2}},\bar{H}_{\bar{3}} \right\}
\end{equation}
Flipped $SU(5)$ model building has two very nice features which
are generally not found in typical unified models: (i) a natural
solution to the doublet ($H_2$)-triplet($H_3$) splitting problem
of the electroweak Higgs pentaplets $h,\bar{h}$ through the
trilinear coupling of the Higgs fields: $H_{\bf{10}} \cdot
H_{\bf{10}} \cdot h_{\bf{5}} \rightarrow \left\langle \nu^c_H
\right\rangle d^c_H H_3$, and (ii) an automatic see-saw mechanism
that provide heavy right-handed neutrino mass through the coupling
to singlet fields $\phi$, $F_{\bf{10}} \cdot {\bar
H}_{\overline{\bf{10}}} \cdot \phi \rightarrow \left\langle
\nu^c_{\bar{H}}\right\rangle \nu^c \phi$.

The generic superpotential $W$ for a flipped $SU(5)$ model will be
of the form :
\begin{equation}
\lambda_1 FFh+\lambda_2 F\bar{f}\bar{h}+\lambda_3 \bar{f}l^c h+
\lambda_4 F\bar{H}\phi +\lambda_5 HHh+\lambda_6
\bar{H}\bar{H}\bar{h}+ \cdots\in W  \label{sp}
\end{equation}
the first three terms provide masses for the quarks and leptons,
the fourth is responsible for the heavy right-handed neutrino mass
and the last two terms are responsible for the doublet-triplet
splitting mechanism \cite{AEHN}.

\section{Some Models with Fluxes}

\subsection{$N_{flux}=192$}

The most ideal situation is to preserve supersymmetry both in the
closed string and open string sectors in the spirit of this flux
construction.  However we found that it is difficult to achieve.
An example of this is shown in Table \ref{192}.  Although this
example is supersymmetric both in the open and closed string
sectors, satisfies the conditions for cancellation of RR charges,
and yields a three generation flipped $SU(5)$ model with a
complete but extended Higgs sector, it does not satisfy the
K-theory constraints.

\begin{table}[h]
\begin{center}
\footnotesize
\begin{tabular}{|@{}c@{}|c||@{}c@{}c@{}c@{}||c@{}|c@{}||c|c|c
|c|@{}c@{}|@{}c@{}|@{}c@{}|@{}c@{}|@{}c@{}|@{}c@{}|@{}c@{}|}
\hline

stk & $N$ & ($n_1$, $m_1$) & ($n_2$, $m_2$) & ($n_3$, $m_3$) & A &
S & $b$ & $b'$ & $c$ & $c'$ & $d$ & $d'$ & $e$ & $e'$ & $f$ & $f'$
& $D7_2$
  \\ \hline \hline

$a$ & 10 & ( 1, 0) & (-1,-1) & (-2, 1) & 2 & -2 & -12 & 24 & 1 &
-3 & 1 & -3 & 0(1) & -2 & 0(1) & -2 & 2   \\ \hline

$b$ & 2 & ( 3,-1) & (-5, 1) & ( 4,-1) & 332 & 148 & - & - & 7 & 15
& 7 & 15 & 12 & 16 & 12 & 16 & 12    \\ \hline

$c$ & 2 & (-2, 1) & ( 2, 1) & (-1, 0) & 0 & 0 & - & - & - & - &
0(0) & 0(16) & 0(0) & 0(9) & 0(0) & 0(9) & 2   \\ \hline

$d$ & 2 & (-2, 1) & ( 2, 1) & (-1, 0) & 0 & 0 & - & - & - & - & -
& - & 0(0) & 0(9) & 0(0) & 0(9) & 2   \\ \hline

$e$ & 2 & (-1, 1) & ( 1, 1) & (-1, 0) & 0 & 0 & - & - & - & - & -
& - & - & - & 0(0) & 0(4) & 1  \\ \hline

$f$ & 2 & (-1, 1) & ( 1, 1) & (-1, 0) & 0 & 0 & - & - & - & - & -
& - & - & - & - & - & 1  \\ \hline

$O7_2$ & 6 & ( 0, 1) & ( 1, 0) & ( 0,-1) & - & - & - & - & - & -
& - & - & - & - & - & - & -   \\
\hline

\end{tabular}
\caption{List of wrapping numbers and intersection numbers for
three-fluxes $N_{flux}=192$. The number in parenthesis indicates
the multiplicity of non-chiral pairs. Here $x_A=62$, $x_B=1$,
$x_C=1$, and $x_D=2$.  It is obvious that the first K-theory
constraint is not satisfied.  The gauge symmetry is $U(5)\times
U(1)^5 \times USp(6)$.}  \label{192}
\end{center}
\end{table}

\subsection{$N_{flux}=128$}

We present an example for $N_{flux}=128$ with four stacks of
magnetized D-branes as well as two filler branes presented in
Table \ref{128}.  Although this particular model does not contain
flipped $SU(5)$ symmetry, it is a consistent solution of the RR
tadpole conditions and the K-theory constraints, and is
supersymmetric both in the open and closed string sectors.  The
gauge symmetry is

\begin{equation}
U(5)\times U(1)\times USp(4) \times USp(4)
\end{equation}

\begin{table}[h]
\begin{center}
\footnotesize
\begin{tabular}{|@{}c@{}|c||@{}c@{}c@{}c@{}||c@{}|c@{}||c|c|c|c|}
\hline

stk & $N$ & ($n_1$, $m_1$) & ($n_2$, $m_2$) & ($n_3$, $m_3$) & A &
S & $b$ & $b'$ & $D3$ & $D7_2$
  \\ \hline \hline

$a$ & 10 & ( 1, 0) & (-1,-1) & (-2, 1) & 2 & -2 & -16 & 24 & 0(1)
& 2  \\ \hline

$b$ & 2 & ( 3,-2) & (-3, 1) & ( 4,-1) & 374 & 202 & - & - & -2 &
12
\\ \hline

$O3$ & 4 & ( 1, 0) & ( 1, 0) & ( 1, 0) & - & - & - & - & - & -
\\ \hline

$O7_2$ & 4 & ( 0, 1) & ( 1, 0) & ( 0,-1) & - & - & - & - & - & -
\\ \hline

\end{tabular}
\caption{$ N_{flux}=128$.  The number stacks is only two plus two
filler branes, though it has very few exotic particles, we have
too few stacks to complete the cancellation of $U(1)_X$ mass. Here
$x_A=27$, $x_B=1$, $x_C=1$, and $x_D=2$.} \label{128}
\end{center}
\end{table}

\subsection{$N_{flux}=1\times 64$}

\begin{table}[h]
\begin{center}
\footnotesize
\begin{tabular}{|@{}c@{}|c||@{}c@{}c@{}c@{}||c|c||c|c|
@{}c@{}|c|@{}c@{}|@{}c@{}|@{}c@{}|@{}c@{}|@{}c@{}|@{}c@{}|} \hline

stk & $N$ & ($n_1$, $m_1$) & ($n_2$, $m_2$) & ($n_3$, $m_3$) & A &
S & $b$ & $b'$ & $c$ & $c'$ & $d$ & $d'$ & $e$ & $e'$ & $f$ & $f'$
  \\ \hline \hline

$a$ & 10 & ( 1, 0) & (-1,-1) & (-2, 1) & 2 & -2 & -8 & 12 & -8 &
12 & 0(0) & 0(8) & 0(1) & 4 & 0(1) & 4   \\ \hline

$b$ & 2 & ( 1,-1) & (-3, 1) & ( 4,-1) & 84 & 12 & - & - & 0(0) &
96 & 8 & 12 & 4 & 0(6) & 4 & 0(6)   \\ \hline

$c$ & 2 & ( 1,-1) & (-3, 1) & ( 4,-1) & 84 & 12 & - & - & - & - &
8 & 12 & 4 & 0(6) & 4 & 0(6)   \\ \hline

$d$ & 2 & (-1, 0) & ( 1, 1) & (-2, 1) & 2 & -2 & - & - & - & - & -
& - & 0(1) & 4 & 0(1) & 4   \\ \hline

$e$ & 2 & ( 1, 1) & ( 1, 0) & ( 2,-1) & 2 & -2 & - & - & - & - & -
& - & - & - & 0(0) & 8
\\ \hline

$f$ & 2 & ( 1, 1) & ( 1, 0) & ( 2,-1) & 2 & -2 & - & - & - & - & - & - & - & - & - & -   \\
\hline

\end{tabular}
\caption{List of intersection numbers for $N_{flux}=64$ with gauge
group $U(5)\times U(1)^5$. The number in parenthesis indicates the
multiplicity of non-chiral pairs.} \label{64}
\end{center}
\end{table}

In this example, we use two sets of parallel D-branes and all
conditions are satisfied.  No filler brane is needed, and
$x_A=22$, $x_B=1$, $x_C=1$, and $x_D=2$.  The complete $(n^i_a,
m^i_a)$ and $SU(5)\times U(1)_X$ spectrum are listed in Table
\ref{64} and \ref{64S}, and $U(1)_X$ is
\begin{equation}
U(1)_X=\frac{1}{2}(U(1)_a-5U(1)_b+5U(1)_c-5U(1)_d+5U(1)_e-5U(1)_f)
\end{equation}

The four global $U(1)$s from the Green-Schwarz mechanism are given
respectively:
\begin{eqnarray}
 U(1)_1 & = & 24U(1)_b+24U(1)_c+4U(1)_e+4U(1)_f  \nonumber\\
 U(1)_2 & = & 20U(1)_a+8U(1)_b+8U(1)_c+4U(1)_d   \nonumber\\
 U(1)_3 & = & -10U(1)_a+6U(1)_b+6U(1)_c-2U(1)_d-2U(1)_e-2U(1)_f   \nonumber\\
 U(1)_4 & = & -2U(1)_b-2U(1)_c
\end{eqnarray}

From Table \ref{64S} we found that none of the global $U(1)$s from
the G-S anomaly cancellation mechanism provides Yukawa couplings
required for generation of mass terms in superpotential
(\ref{sp}).  However, $U(1)_X$ admits these Yukawa couplings, and
if we require the other anomaly-free and massless combination
$U(1)_Y$ does as well, two conditions can be considered.  The
first one is to demand all the Yukawa couplings from the assigned
intersections, and an example of the $U(1)_Y$ and the
corresponding combinations of representations are listed as
follows:
\begin{equation}
U(1)_Y^1 = 5U(1)_a-25U(1)_b+25U(1)_c-25U(1)_d-38U(1)_e+38U(1)_f
\end{equation}
\begin{eqnarray}
F F h \; &\rightarrow &\;(\mathbf{10},\mathbf{1})
(\mathbf{10},\mathbf{1}) (\mathbf{5}_a,\mathbf{1}_d)^\star \nonumber \\
F \bar{f} \bar{h} \; &\rightarrow &\; (\mathbf{10},\mathbf{1})
(\mathbf{\overline{5}}_a,\mathbf{1}_b)
(\mathbf{\overline{5}}_a,\mathbf{\overline{1}}_d)^\star \nonumber \\
\bar{f} l^c h \; &\rightarrow &\;
(\mathbf{\overline{5}}_a,\mathbf{1}_b)
(\mathbf{1}_c,\mathbf{\overline{1}}_d)
(\mathbf{5}_a,\mathbf{1}_d)^\star
\nonumber \\
F \bar{H} \phi \; &\rightarrow &\; (\mathbf{10},\mathbf{1})
(\overline{\mathbf{10}},\mathbf{1}) (\mathbf{1}_b,\mathbf{1}_c)
\nonumber \\
H H h \; &\rightarrow &\; (\mathbf{10},\mathbf{1})
(\mathbf{10},\mathbf{1}) (\mathbf{5}_a,\mathbf{1}_d)^\star \nonumber \\
\bar{H} \bar{H} \bar{h} \; &\rightarrow &\;
(\overline{\mathbf{10}},\mathbf{1})(\overline{\mathbf{10}},\mathbf{1})
(\mathbf{\overline{5}}_a,\mathbf{\overline{1}}_d)^\star
\end{eqnarray}

If we do not require the Higgs pentaplet $\bar{h}'$ coupled with
the chiral fermions in the term $F \bar{f} \bar{h}'$ to be the
same as the Higgs pentaplet $\bar{h}$ coupled to $\bar{H}$, then
we expect a mixture state $\bar{h}_x = c \bar{h}' + s \bar{h}$ of
these two different Higgs pentaplets in the Higgs sector,
therefore
\begin{equation}
U(1)_Y^2 = U(1)_b-U(1)_c+U(1)_e-U(1)_f
\end{equation}
\begin{eqnarray}
F \bar{f} \bar{h}' \; &\rightarrow &\; (\mathbf{10},\mathbf{1})
(\mathbf{\overline{5}}_a,\mathbf{1}_b)
(\mathbf{\overline{5}}_a,\mathbf{1}_c) \nonumber \\
\bar{H} \bar{H} \bar{h} \; &\rightarrow &\;
(\overline{\mathbf{10}},\mathbf{1})(\overline{\mathbf{10}},\mathbf{1})
(\mathbf{\overline{5}}_a,\mathbf{\overline{1}}_d)^\star
\end{eqnarray}

We should also notice that the superfluous
$\bar{\mathbf{5}},\mathbf{5}$, and $\overline{\mathbf{10}}$ representations may be
ostracized from the low energy spectrum through trilinear
couplings of the generic form
$\bar{\mathbf{5}}\cdot\mathbf{5}\cdot\mathbf{1}$ and
$\overline{\mathbf{10}}\cdot\mathbf{10}\cdot\mathbf{1}$ satisfying the
gauged $U(1)$ symmetries, where the singlets are assumed to
acquire string scale \textit{vev}s.

\begin{table}[h]
\begin{center}
\footnotesize
\begin{tabular}{|@{}c@{}||@{}c@{}||@{}c@{}|@{}c@{}|@{}c@{}|@{}c@{}|@{}c@{}|@{}c@{}||@{}c@{}||@{}c@{}|
@{}c@{}|@{}c@{}|@{}c@{}||@{}c@{}|@{}c@{}|} \hline

 Rep. & Multi. &$U(1)_a$&$U(1)_b$&$U(1)_c$& $U(1)_d$
& $U(1)_e$ & $U(1)_f$ & $U(1)_X$ & $U(1)_1$ & $U(1)_2$ &
$U(1)_3$ & $U(1)_4$ & $U(1)_Y^1$ & $U(1)_Y^2$ \\
\hline \hline

$(10,1)$ & 3 & 2 & 0 & 0 & 0 & 0 & 0 & 1 & 0 & 40 & -20 & 0 & 10 & 0 \\

$(\bar{5}_a ,1_b)$ & 3 & -1 & 1 & 0 & 0  & 0 & 0 & -3 & 24 & -12 &
16 & -2 & -30 & 1  \\

$(1_c ,\bar{1}_d)$ & 3 & 0 & 0 & 1 & -1 & 0 & 0 & 5 & 24 & 4 & 8 &
-2 & 50 & -1
\\ \hline \hline

$(10,1)$ & 1 & 2 & 0 & 0 & 0 & 0 & 0 & 1 & 0 & 40 & -20 & 0 & 10 &
0 \\

$(\overline{10},1)$ & 1 & -2 & 0 & 0 & 0 & 0 & 0 & -1 & 0 & -40 &
20 & 0 & -10 & 0  \\
\hline

$(5_a,1_d)^\star$ & 1 & 1 & 0 & 0 & 1 & 0 & 0 & -2 & 0 & 24 & -12
& 0 & -20 & 0   \\

$^1(\bar{5}_a ,\bar{1}_d)^\star$/$^2\bar{h}_x$ & 1 & $^1$-1 &
$^1$0 & $^1$0 & $^1$-1 & $^1$0 & $^1$0 & 2 & $^1$0 & $^1$-24 &
$^1$12 & $^1$0 & $^1$20 & $^2$-1/ 0
\\ \hline

$(1_b ,1_c)$ & 4 & 0 & 1 & 1 & 0 & 0 & 0 & 0 & 48 & 16 & 12 & -4 &
0 & 0  \\  \hline \hline

$(\overline{15},1)$ & 2 & -2 & 0 & 0 & 0 & 0 & 0 & -1 & 0 & -40 &
20 & 0 & -10 & 0   \\

$(\overline{10},1)$ & 1 & -2 & 0 & 0 & 0 & 0 & 0 & -1 & 0 & -40 &
20 & 0 & -10 & 0   \\
\hline

$(\bar{5}_a ,1_b)$ & 5 & -1 & 1 & 0 & 0 & 0 & 0 & -3 & 24 & -12 &
16 & -2 & -30 & 1   \\

$(5_a ,1_b)$ & 12 & 1 & 1 & 0 & 0 & 0 & 0 & -2 & 24 & 28 & -4 & -2
& -20 & 1   \\

$(\bar{5}_a ,1_c)$ & 8 & -1 & 0 & 1 & 0 & 0 & 0 & 2 & 24 & -12 &
16 & -2 & 20 & -1   \\

$(5_a ,1_c)$ & 12 & 1 & 0 & 1 & 0 & 0 & 0 & 3 & 24 & 28 & -4 & -2
& 30 & -1   \\

$(5_a ,1_e)$ & 4 & 1 & 0 & 0 & 0 & 1 & 0 & 3 & 4 & 20 & -12 & 0 &
-33 & 1  \\

$(5_a ,1_f)$ & 4 & 1 & 0 & 0 & 0 & 0 & 1 & -2 & 4 & 20 & -12 & 0 &
43 & -1  \\ \hline

$(1_b ,1_c)$ & 92 & 0 & 1 & 1 & 0 & 0 & 0 & 0 & 48 & 16 & 12 & -4
& 0 & 0   \\

$(1_b ,\bar{1}_d)$ & 8 & 0 & 1 & 0 & -1 & 0 & 0 & 0 & 24 & 4 & 8
& -2 & 0 & 1   \\

$(1_b ,1_d)$ & 12 & 0 & 1 & 0 & 1 & 0 & 0 & -5 & 24 & 12 & 4 & -2
& -50 & 1   \\

$(1_b ,\bar{1}_e)$ & 4 & 0 & 1 & 0 & 0 & -1 & 0 & -5 & 20 & 8 & 8
& -2 & 13 & 0   \\

$(1_b ,\bar{1}_f)$ & 4 & 0 & 1 & 0 & 0 & 0 & -1 & 0 & 8 & 8 & -2
& 0 & -63 & 2   \\

$(1_c ,\bar{1}_d)$ & 5 & 0 & 0 & 1 & -1 & 0 & 0 & 5 & 24 & 4 & 8
& -2 & 50 & -1   \\

$(1_c ,1_d)$ & 12 & 0 & 0 & 1 & 1 & 0 & 0 & 0 & 24 & 12 & 4
& -2 & 0 & -1   \\

$(1_c ,\bar{1}_e)$ & 4 & 0 & 0 & 1 & 0 & -1 & 0 & 0 & 20 & 8 & 8
& -2 & 63 & -2   \\

$(1_c ,\bar{1}_f)$ & 4 & 0 & 0 & 1 & 0 & 0 & -1 & 5 & 20 & 8 & 8
& -2 & -13 & 0   \\

$(1d ,1_e)$ & 4 & 0 & 0 & 0 & 1 & 1 & 0 & 0 & 4 & 4 & -4 & 0 & -63
& 1  \\

$(1d ,1_f)$ & 4 & 0 & 0 & 0 & 1 & 0 & 1 & -5 & 4 & 4 & -4 & 0 &
13 & -1  \\

$(1,1)$ & 12 & 0 & 2 & 0 & 0 & 0 & 0 & -5 & 48 & 16 & 12 & -4 &
-50 & 2   \\

$(1,1)$ & 12 & 0 & 0 & 2 & 0 & 0 & 0 & 5 & 48 & 16 & 12 & -4 & 50
& -2   \\

$(\bar{1},\bar{1})$ & 2 & 0 & 0 & 0 & -2 & 0 & 0 & 5 & 0 & -8 & 4
& 0 & 50 & 0   \\

$(\bar{1},\bar{1})$ & 2 & 0 & 0 & 0 & 0 & -2 & 0 & -5 & -8 & 0 & 4
& 0 & 76 & -2   \\

$(\bar{1},\bar{1})$ & 2 & 0 & 0 & 0 & 0 & 0 & -2 & 5 & -8 & 0 & 4
& 0 & -76 & 2   \\
\hline

$(5_a,1_d)^\star$ & 7 & 1 & 0 & 0 & 1 & 0 & 0 & -2 & 0 & 24 & -12
& 0 & -20 & 0   \\

$(\bar{5}_a ,\bar{1}_d)^\star$ & 7 & -1 & 0 & 0 & -1 & 0 & 0 & 2 &
0 & -24 & 12 & 0 & 20 & 0  \\
\hline

\multicolumn{15}{|c|}{Additional non-chiral Matter}\\ \hline

\end{tabular}
\caption{The spectrum of $U(5)\times U(1)^5$, or $SU(5)\times
U(1)_X\times U(1)_Y$, with the four global $U(1)$s from the
Green-Schwarz mechanism.  The $\star'd$ representations indicate
vector-like matter.  We list the two cases for the $U(1)_Y$.}
\label{64S}
\end{center}
\end{table}

\newpage

\section{Conclusions}

In this paper, we built flipped $SU(5)$ GUT models
using D-brane constructions on a Type IIB $\mathbf{T^6} /(\Z_2
\times \Z_2)$ orientifold with supergravity fluxes turned on.  We
considered both supersymmetric and non-supersymmetric fluxes in
the closed string sector, and we claim that only the
non-supersymmetric (soft-breaking) cases of flipped $SU(5)$ we
have found are consistent with all the constraints of string
theory including K-theory and supersymmetry in the open string
sector.

The model that we have presented in Table \ref{64S} contains
three-generations of chiral fermions and a complete GUT and
electroweak Higgs sector.  It also includes extra matter such as
two copies of the symmetric representation of $SU(5)$ as well as
many extra bi-fundamental and vector-like representations, which
result from the large D9-brane co-prime numbers $(n_a^i,
m_a^i)_{D9_a}$ needed for the required compensation of the induced
three-form flux contributions to the D3 RR charge.

As mentioned above, the non-supersymmetric flux ($N_{flux}=64$) in
this particular flipped $SU(5)$ model breaks supersymmetry in the
closed string sector.  This leads to a mechanism of soft
supersymmetry breaking at a mass scale $M_{soft}\sim
\frac{M_{string}^2}{M_{Pl}}$ which implies an intermediate string
scale or an inhomogeneous warp factor in the internal space to
stabilize the electroweak scale \cite{Camara:2003Soft,
Lust:2004fi, CveticLiLiu}.  With this non-supersymmetric flux
present, soft supersymmetry breaking terms may be manifested in
the effective action of open string fields.  Detailed studies in
soft-breaking mechanism and some trial investigations into the
effective low energy scenario were studied in \cite{Lust:2004fi,
Kane:2004hm}. Combined with a Yukawa coupling analysis
\cite{Cremades:2003qj}, this may provide a clear picture of the
low energy physics which we defer for future work.

The four global $U(1)$ symmetries from the G-S anomaly
cancellation forbid all the Yukawa couplings necessary for the
generation of quark and lepton masses, although if we ignore these
global $U(1)$ factors and focus only on the $U(1)_X$ and $U(1)_Y$
symmetries, then we find that all of the required Yukawa couplings
in (\ref{sp}) are present, as well as those needed for making the
extra matter in the model obtain mass $\mathcal{O}(M_{string})$.
We need to keep in mind that global $U(1)$ symmetries are valid to
all orders in perturbation theory, and can be broken by
non-perturbative instanton effects \cite{Kachru:Uranga}. To solve
this problem without these instanton effects, one possibility one
may entertain is to use singlets, suitably charged, to trigger
spontaneous breaking of global $U(1)$s as well as of the local
$U(1)_Y$ at the string scale, while leaving $U(1)_X$ intact.  In
the case of global $U(1)$s one may hope that we will end up with
invisible axion-like bosons. The interested reader may check from
Table \ref{64S} that such singlets with appropriate charges do
exist. Another possibility is that we may need a new D-brane
configuration. It has been recently shown that the RR, NSNS and
metric fluxes could contribute negative D6-brane charges in the
Type IIA orientifold with flux compactifications, and thus relax
the RR tadpole cancellation conditions~\cite{Camara:2005dc}, which
is a good basis for future work as well as providing a solution to
the problem of finding a compatible set of global $U(1)$s on
$\mathbf{T^6} /(\Z_2 \times \Z_2)$ orientifold with the Yukawa
couplings.

\section{Acknowledgements}

We thank George V. Kraniotis for his contribution to this work in
its early stages.  The work of D. V. N. is supported by DOE grant
DE-FG03-95-ER-40917.

\end{document}